\documentclass[aps,prl,preprint,superscriptaddress]{revtex4-1}

\usepackage{epsfig}
\usepackage{graphics}
\usepackage{amsfonts}
\usepackage{mathrsfs}
\usepackage{amsmath}
\usepackage{color}
\usepackage{natbib}

\usepackage{textcomp}
\usepackage{graphicx}
\usepackage{bm}
\usepackage{amssymb}
\usepackage{ulem}
\usepackage{xspace}
\usepackage{epstopdf}
\usepackage{dcolumn}
\usepackage{subfigure}
\usepackage{multirow}
\usepackage{diagbox}
\usepackage{array}
\usepackage{geometry}

\usepackage{makecell}
\usepackage[colorlinks=true, letterpaper=true, pdfstartview=FitV, linkcolor=blue, citecolor=blue, urlcolor=blue]{hyperref}

\begin{document}

\title{Defects Engineering of ZrTe$_5$ for Stabilizing Ideal Topological States}

\author{Chia-Hsiu Hsu}
\altaffiliation{These authors contributed equally to this work.}
\affiliation{Division of Physics and Applied Physics, School of Physical and Mathematical Sciences, Nanyang Technological University, 637371, Singapore}
\affiliation{Quantum Materials Science Unit, Okinawa Institute of Science and Technology, Okinawa 904-0495, Japan}
\author{Zezhi Wang}
\altaffiliation{These authors contributed equally to this work.}
\affiliation{Beijing National Laboratory for Condensed Matter Physics, Institute of Physics, Chinese Academy of Sciences, Beijing 100190, China}
\affiliation{School of Physical Sciences, University of Chinese Academy of Sciences, Beijing 100049, China}
\author{Sen Shao}
\altaffiliation{These authors contributed equally to this work.}
\affiliation{Division of Physics and Applied Physics, School of Physical and Mathematical Sciences, Nanyang Technological University, 637371, Singapore}
\author{Yoshinori Okada}
\affiliation{Quantum Materials Science Unit, Okinawa Institute of Science and Technology, Okinawa 904-0495, Japan}
\author{Feng-Chuan Chuang} 
\affiliation{Department of Physics, National Sun Yat-Sen University, Kaohsiung 804, Taiwan}
\author{Dong Xing}
\affiliation{Beijing National Laboratory for Condensed Matter Physics, Institute of Physics, Chinese Academy of Sciences, Beijing 100190, China}
\affiliation{School of Physical Sciences, University of Chinese Academy of Sciences, Beijing 100049, China}
\author{Ilya Belopolski}
\affiliation{School of Electrical and Electronic Engineering, Nanyang Technological University, 637371, Singapore}
\author{Cheng-Long Zhang}
\email{chenglong.zhang@iphy.ac.cn}
\affiliation{Beijing National Laboratory for Condensed Matter Physics, Institute of Physics, Chinese Academy of Sciences, Beijing 100190, China}
\author{Guoqing Chang}
\email{guoqing.chang@ntu.edu.sg}
\affiliation{Division of Physics and Applied Physics, School of Physical and Mathematical Sciences, Nanyang Technological University, 637371, Singapore}


\maketitle
\section{Keywords} Topological Insulators, Defect, First-principles calculations.


\section{Abstract}

  ZrTe$_5$ is a highly tunable, high-mobility topological material that hosts a rich variety of quantum phenomena, making it a promising platform for next-generation quantum technologies. Despite intensive research efforts, experimental studies have reported inconsistent and sometimes conflicting results for its electronic and topological states, largely due to variations in sample quality. Here, through systematic first-principles investigations of all intrinsic point defects, we identify a practical route to achieving stable and ideal topological characteristics in ZrTe$_5$.  We show that the competition between two dominant charged defects, donor-like Zr interstitials and acceptor-like Te vacancies, governs the Fermi-level position. Furthermore, variations in defect density determine the topological phases of the samples. We theoretically propose and experimentally confirm that increasing the Te/Zr ratio during crystal growth effectively suppresses intrinsic defects and stabilizes ZrTe$_5$ in a nearly ideal weak topological insulator state. These findings provide clear guidance for defect control and sample optimization, paving the way toward robust and reproducible realization of topological quantum states in ZrTe$_5$ for future quantum applications.


\section{Introduction}

Topological insulators (TIs) have drawn considerable attention owing to their potential in realizing low-dissipation spintronic and quantum information devices, which exploit their insulating bulk states and topologically protected boundary states \cite{Intro1,Intro2,Intro3,Intro4}. In practice, however, intrinsic defects often introduce unintended bulk conduction, thereby obscuring the distinctive quantum phenomena associated with the boundary \cite{Th_Imp1,Th_Imp2,Th_Imp4,Th_Imp5,Th_Imp6}. Consequently, controlling impurities to approach the ideal TI limit has become a central challenge in materials synthesis and has long been a key experimental focus \cite{Intro1,Ex_Imp1,Ex_Imp3}. A prototypical example is the Bi$_2$Se$_3$ family \cite{Bi2Te31}, where intrinsic defects typically drive the Fermi level deep into the conduction band \cite{D_Bi2Te31,D_Bi2Te32,D_Bi2Te33}. Various experimental approaches, such as chemical doping \cite{Chdop1,Chdop2,Chdop3,Chdop4,Chdop5,Chdop6,Chdop7} and electrostatic gating \cite{Egate1, Egate2}, have been employed to suppress defect-induced bulk conduction, successfully yielding samples with genuinely insulating bulk behavior in the Bi$_2$Se$_3$ family.

In contrast, for ZrTe$_5$ \cite{PRX,StrainZrTe5}, effective defect control to stabilize a clean topological insulating phase remains elusive. ZrTe$_5$ exhibits exceptionally high carrier mobility and exotic optical and transport responses, positioning it as a promising platform for next-generation quantum technologies \cite{mobility,optical,Magneto,AHE,mono,FL6,FL7}. Yet, its electronic and topological character is extremely sensitive to growth conditions, with experimentally reported phases ranging from (i) electron-doped weak TIs (WTIs) \cite{cvt9,cvt10,cvt11,FL7}, (ii) electron-doped strong TIs (STIs) \cite{magspec1,magspec2,cvt13}, (iii) hole-doped STIs \cite{cvt6,cvt8}, to (iv) hole-doped WTIs \cite{cvt5,cvt7,FL6} [Figure \ref{fig1}(a)]. Such highly sample-dependent and even self-contradictory results have prompted the community to reconsider the dominant role of impurities and intrinsic defects in defining the electronic ground state of ZrTe$_5$. Combined spectroscopic and transport studies, supported by theoretical calculations, indicate that variations in impurity and defect populations account for the distinct behaviors observed between chemical-vapor-transport (CVT)-grown and flux-grown crystals \cite{cvt12,TempC,TempD,TempEn,cvt14,TempF}.  Despite the intensive research interests, an effective defect management strategy during crystal growth has yet to be established.\\


\section{Results and discussion}
\begin{figure}[htbb]
	\centering
	\includegraphics[width=1\textwidth]{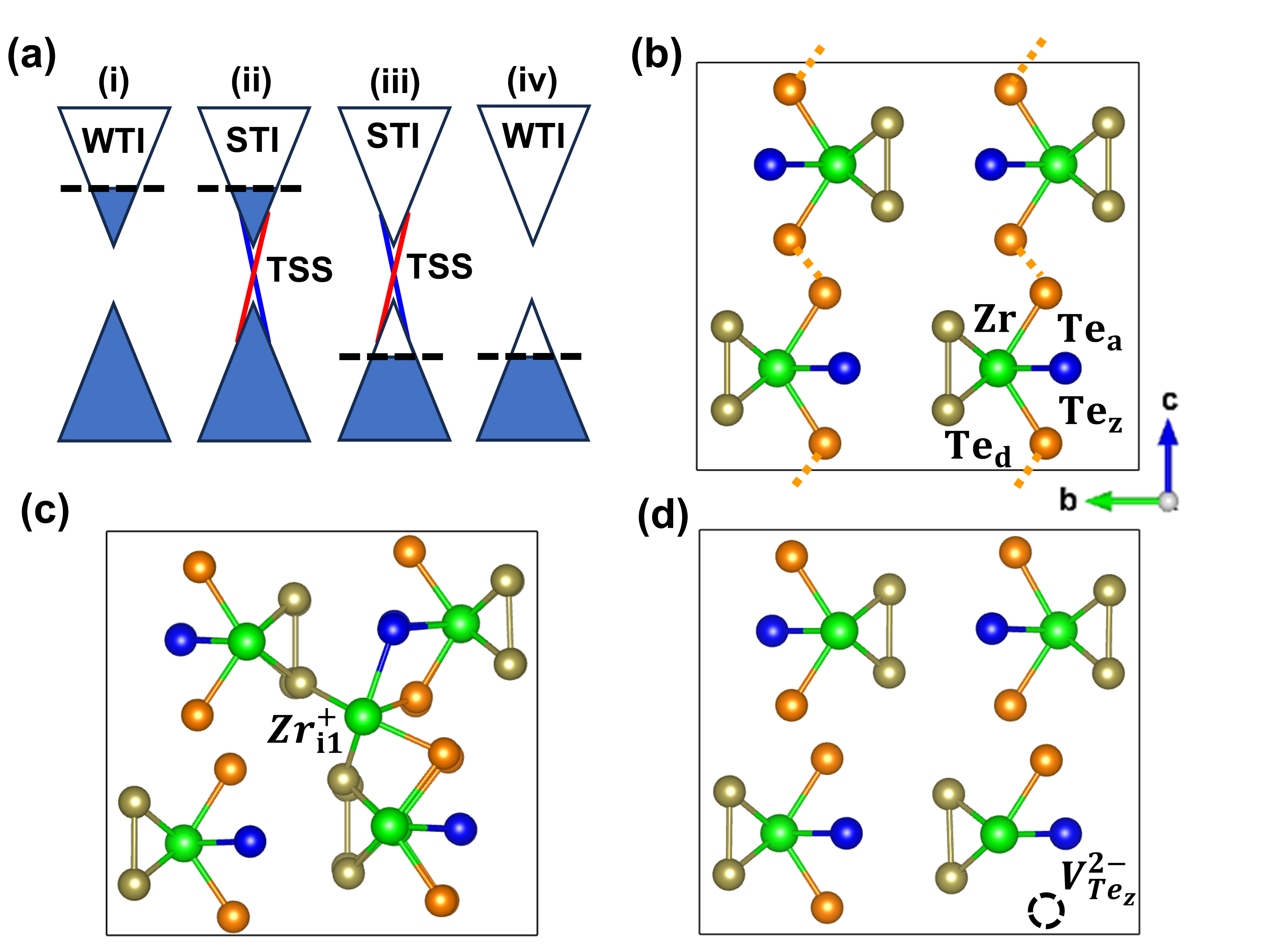}
	\begin{center}
		\caption{(a)  Reported phase:  (i) electron-doped WTIs, (ii) electron-doped STIs, (iii) hole-doped STIs and (iv) hole-doped WTIs. (b) Crystal structure of ZrTe$_{5}$. Green, blue, brown and orange spheres represent Zr, Te$_z$, Te$_d$ amd Te$_a$, respectively. Schematic diagram of structure with (c) $\mathrm{Zr}_{\mathrm{i1}}^{+}$ defect and (d) $\mathrm{V}_{\mathrm{Te}_{\mathrm{z}}}^{2-}$ defect.}
		\label{fig1}
	\end{center}
\end{figure}

To overcome this limitation, we perform a comprehensive study on the intrinsic point defects in ZrTe$_{5}$, which can be categorized into three types: vacancies ($\mathrm{V}_{\mathrm{X}}$), interstitials  ($\mathrm{X}_{\mathrm{i}}$), and antisite defects ($\mathrm{X}_{\mathrm{Y}}$), where $\mathrm{V}$ and $\mathrm{i}$ indicate a vacancy and an interstitial site, respectively, and $\mathrm{X}$ and $\mathrm{Y}$ represent Zr or Te atoms. According to the crystal symmetry of ZrTe$_{5}$ (space group 63), there is one symmetry-inequivalent Zr site and three inequivalent Te sites [Figure \ref{fig1}(b)]. These Te sites can be categorized as the apical site ($\mathrm{Te}_{\mathrm{a}}$), the dimer site ($\mathrm{Te}_{\mathrm{d}}$) forming the prismatic ZrTe$_3$ chains along the $a$ axis, and the zigzag site ($\mathrm{Te}_{\mathrm{z}}$) that connects adjacent ZrTe$_3$ chains along the $c$ axis to complete the ZrTe$_5$ layers in the $a$–$c$ plane. Therefore, there are three distinct Te vacancies ($\mathrm{V}_{\mathrm{Te}_{\mathrm{d}}}$, $\mathrm{V}_{\mathrm{Te}_{\mathrm{z}}}$, and $\mathrm{V}_{\mathrm{Te}_{\mathrm{a}}}$), one possible Zr vacancy ($\mathrm{V}_{\mathrm{Zr}}$), three antisite defects where Zr substitutes Te ($\mathrm{Zr}_{\mathrm{Te}_{\mathrm{d}}}$, $\mathrm{Zr}_{\mathrm{Te}_{\mathrm{z}}}$, and $\mathrm{Zr}_{\mathrm{Te}_{\mathrm{a}}}$), and one antisite defect where Te replaces Zr ($\mathrm{Te}_{\mathrm{Zr}}$) (Figure S1).  In addition, inserting a Te atom into the van der Waals gap along the $ac$ plane yields nine distinct interstitial configurations (Figure S2). Similarly, Zr intercalation also produces another nine interstitial configurations (Figure S3). 


 We first examine the influence of intrinsic defects on the Fermi level of ZrTe$_5$. By calculating the defect formation energies as a function of the Fermi level relative to the valence-band maximum, we identify the stable charge states of each defect [see Supporting Information method and Figure S4 for details). Among all the defects, we identified that $\mathrm{Zr}_{\mathrm{i1}}^{+}$ [Figure~\ref{fig1}(c)] /$\mathrm{Zr}_{\mathrm{i8}}^{+}$ and $\mathrm{V}_{\mathrm{Te}_{\mathrm{z}}}^{2-}$ are the primary charged defects responsible for modifying the Fermi level compared to the ideal crystal. The charge states of each defect are indicated by the superscripts in their notations. Specifically, interstitial Zr acts as the dominant donor that shifts the Fermi level upward, whereas the Te vacancy at the zigzag site serves as the main acceptor. The behavior of interstitial Zr is consistent with the conventional ionic picture, since Zr is a metallic element prone to electron loss, an interstitial Zr atom naturally forms a $\mathrm{Zr}_{\mathrm{i1}}^{+}$ defect. In contrast, the Te vacancy in ZrTe$_5$ deviates markedly from this conventional expectation.

\begin{figure}[ptb]
	\centering
	\includegraphics[width=1.0\textwidth]{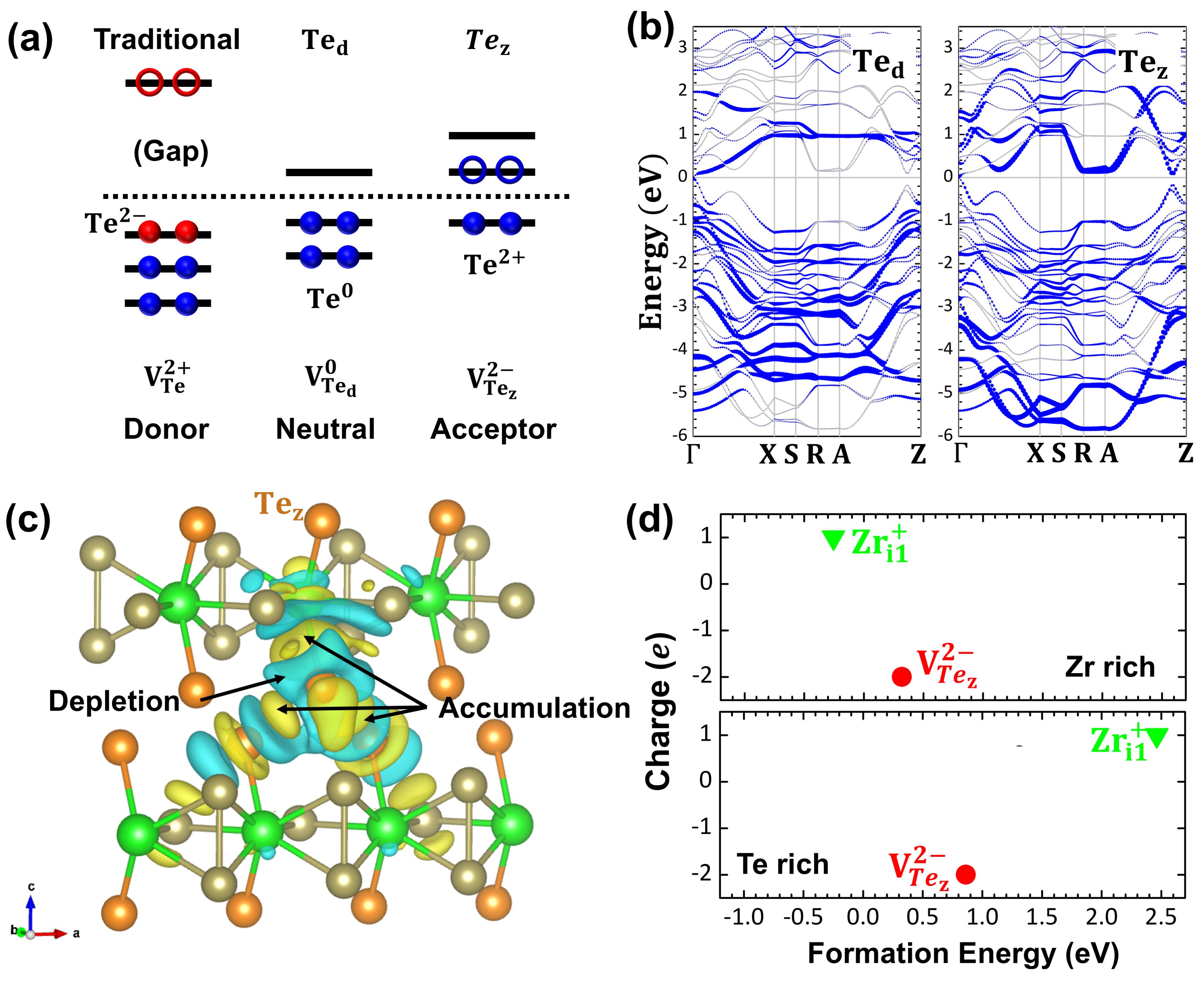}
	\caption{(a) Schematic illustration of donor-like, neutral, and acceptor-like behavior of Te vacancies. (b) Band structure with the $p$ orbital contribution from the $\mathrm{Te}_{\mathrm{d}}$ and $\mathrm{Te}_{\mathrm{z}}$ atoms.(c) Charge density difference for the $\mathrm{Te}_{\mathrm{z}}$ site. Yellow and cyan indicate charge accumulation and depletion, respectively. (d) Formation energies for charged defects under Zr-rich and Te-rich conditions, respectively.} \label{fig2}
\end{figure}

Typically, Te, being highly electronegative, tends to attract electrons. Consequently, removing a Te atom to form a vacancy, $\mathrm{V}_{\mathrm{Te}}^{2+}$ [Figure \ref{fig2}(a), left], would conventionally create a donor-type defect. This discrepancy arises because ZrTe$_5$ is not a conventional large-gap insulator [Figure \ref{fig2}(a), left]; instead, it is a small-gap WTI with Te $p$ orbitals crossing the Fermi level and contributing states to both conduction and valence bands. When there is one $p$ orbital lying above the Fermi level and the other two below the Fermi level, the Te atom is effectively charge neutral. This leads to the neutral vacancy $\mathrm{V}_{\mathrm{Te}_{\mathrm{d}}}^0$ [Figure \ref{fig2}(a), middle]. When two of the three Te $p$ orbitals lie above the Fermi level, a Te atom donates two electrons to the lattice [Figure \ref{fig2}(a), right]; removing it to create a vacancy, therefore, withdraws two electrons, resulting in the $\mathrm{V}_{\mathrm{Te}_{\mathrm{z}}}^{2-}$ defect and a downward shift of the Fermi level.

These trends are confirmed by our first-principles calculations. Figure~\ref{fig2}(b) show the $p$ orbital contributions from the $\mathrm{Te}_{\mathrm{d}}$ and $\mathrm{Te}_{\mathrm{z}}$ sites in primitive ZrTe$_5$, with the relevant states highlighted by blue circles. We could see that the $p$ orbital has large contributions in both conduction and valence bands. What is more, whereas $\mathrm{Te}_{\mathrm{d}}$ is dominated by occupied states, $\mathrm{Te}_{\mathrm{z}}$ exhibits more unoccupied orbitals. To further visualize how these orbital characteristics manifest in real space, we examined the charge density difference associated with the $\mathrm{Te}_{\mathrm{z}}$ site [Figure~\ref{fig2}(c)]. The charge density difference was calculated as the total charge density of the full system minus the charge density of the defective system (without $\mathrm{Te}_{\mathrm{z}}$) and the isolated $\mathrm{Te}_{\mathrm{z}}$ atom. This definition highlights the redistribution of electronic charge upon the incorporation of $\mathrm{Te}_{\mathrm{z}}$, where regions of charge accumulation (yellow) and depletion (cyan) can be identified. It can be observed that the charge density decreases around the $\mathrm{Te}_{\mathrm{z}}$ atom, while the accumulation mainly occurs in the interatomic regions, i.e., between the neighboring Zr and $\mathrm{Te}_{\mathrm{z}}$ atoms. This indicates that the $\mathrm{Te}_{\mathrm{z}}$ atom primarily loses electronic charge, thereby tending to form $\mathrm{V}_{\mathrm{Te}_{\mathrm{z}}}^{2-}$ defect. 

Notably, the relative ratio of $\mathrm{Zr}_{\mathrm{i1}}^{+}$ and $\mathrm{V}_{\mathrm{Te}_{\mathrm{z}}}^{2-}$ defects can be affected by the growth conditions. 
Under Zr-rich conditions, where Zr ratio is much larger than Te (see Supporting Information method section), the formation energy of $\mathrm{Zr}_{\mathrm{i1}}^{+}$ is lower than that of  $\mathrm{V}_{\mathrm{Te}_{\mathrm{z}}}^{2-}$, leading to a higher concentration of  $\mathrm{Zr}_{\mathrm{i1}}^{+}$ defect [Figure \ref{fig2}(d), top]. Conversely, under Te-rich conditions, the relative stability is reversed, and  $\mathrm{V}_{\mathrm{Te}_{\mathrm{z}}}^{2-}$ defect becomes more abundant [Figure \ref{fig2}(d), bottom]. Because $\mathrm{Zr}_{\mathrm{i1}}^{+}$ donors raise the Fermi level while $\mathrm{V}_{\mathrm{Te_{z}}}^{2-}$ acceptors lower it, their relative concentrations directly determine the Fermi-level position and thus modify the resulting Fermi level.

\begin{figure*}[ptb]
	\includegraphics[width=1.0\textwidth]{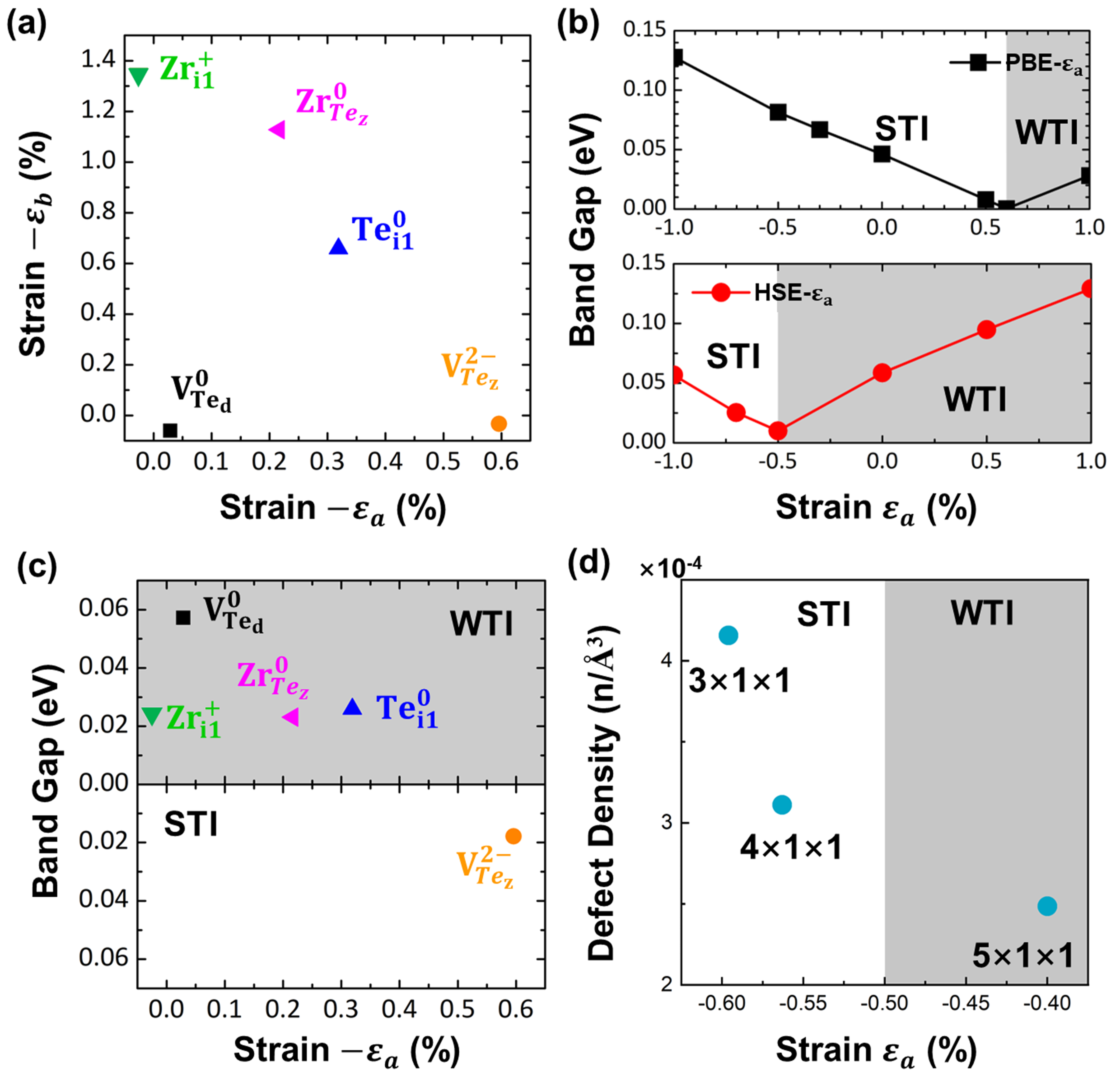}
	\begin{center}
		\caption{ (a) Strain induced by five dominant defects in $ab$ plane. (b) Band Gap as a function of strain $\varepsilon_{a}$ in PBE and HSE calculation, respectively. (c) Band gaps and topological phases of five dominant defects. (d) Strain along a axis and topological phases under various $V_{\mathrm{Te}_{\mathrm{z}}}^{2-}$ defect density (N$_{defect}$/Volume). Calculations are performed with the same number $V_{\mathrm{Te}_{\mathrm{z}}}^{2-}$ defect in 3$\times$1$\times$1, 4$\times$1$\times$1, 5$\times$1$\times$1 supercell, respectively.} \label{fig3}
	\end{center}
\end{figure*}

We next examine the impact of intrinsic defects on the bulk topology of ZrTe$_5$. Motivated by previous studies showing that tensile strain can drive a transition from a strong to a weak topological insulator \cite{FL5}, we investigate how intrinsic defects modify the lattice parameters of ZrTe$_5$ and, consequently, its topological phase. Figure \ref{fig3}(a) summarizes the strain induced by the five dominant defects, including both charged and neutral types. We find that the major defects in ZrTe$_5$ generally compress the lattice, which is consistent with the findings reported for HfTe$_{5}$\cite{TempD}. By applying uniaxial strains along each crystallographic axis, we find that the band gap of ZrTe$_5$ is most sensitive to changes along the $a$ axis and least sensitive to the $c$ axis (Figure S5). Therefore, we use the $a$-axis strain as a representative parameter to illustrate the topological phase transition under strain [Figure \ref{fig3}(b)].

Our PBE calculations show that lattice compression ($\varepsilon < 0$) enlarges the band gap of the STI phase [Figure \ref{fig3}(b), top]. This indicates that intrinsic impurities stabilize the STI phase of ZrTe$_5$, suggesting that only the STI should be experimentally observed. However, this prediction conflicts with experimental findings, where both WTI and STI phases have been reported, and the measured bulk gap is substantially smaller than theoretical predictions \cite{cvt9,cvt10,cvt11,cvt5,cvt7,cvt14,cvt6,cvt8}.

In contrast, our HSE calculations reconcile this discrepancy: ideal ZrTe$_5$ is found to be a WTI [Figure \ref{fig3}(b), bottom]. Under lattice compression ($\varepsilon < 0$) induced by intrinsic defects, the WTI band gap decreases, and once the gap closes and reopens, the system transitions into an STI phase. Figure \ref{fig3}(c) illustrates the defect-induced evolution of the band gap and the corresponding topological character in a 3$\times$1$\times$1 supercell. We note that under the same defect density, $\mathrm{V}_{\mathrm{Te_z}}^{2-}$ is the most pronounced one in driving the topological phase transition of ZrTe$_5$ [Figure \ref{fig3}(c)]. This is because, while most of the defects are compressing the $b$ or $c$ axis, $\mathrm{V}_{\mathrm{Te_z}}^{2-}$  is mainly compressing the $a$ axis. 

We next take $\mathrm{V}_{\mathrm{Te_z}}^{2-}$ as a representative example to demonstrate that the density of intrinsic defects modulates lattice compression, thereby determining the resulting topological phase of the sample. We computed the lattice constants of ZrTe$_5$ under varying $\mathrm{V}_{\mathrm{Te_z}}^{2-}$ defect densities (Table S1).  We found that increasing the defect density drives ZrTe$_5$ into an STI phase [Figure \ref{fig3}(d)].

Bringing together the above insights, our results provide a unified explanation for the emergence of all four distinct phases reported in ZrTe$_5$. As shown in Figure \ref{fig2}, the relative concentrations of different charged defects determine the electron filling in each ZrTe$_5$ sample, while Figure \ref{fig3} reveals that the overall defect density dictates the resulting topological phase. Specifically, Phase (i), an electron-doped WTI, occurs when the overall defect density is low and interstitial Zr dominates as the primary donor-type defect that raises the Fermi level. Phase (ii), an electron-doped STI, appears when the defect density is high and interstitial Zr remains the dominant charged defect. Phase (iii), a hole-doped STI, emerges when the overall defect density is high, but Te vacancies at the zigzag sites prevail as the dominant acceptor-type defects that lower the Fermi level. Finally, Phase (iv), a hole-doped WTI, is realized when the defect density is low and Te vacancies at the zigzag sites are predominant. What is more, our calculations also show that the "finite" defect density will tend to decrease the bulk gap relative to that of ideal crystals [Figure \ref{fig3}(c)]. This is also consistent with the experimental observations, where the gap is much smaller than the predictions in the ideal case.

\begin{figure*}[ptb]
	\includegraphics[width=0.8\textwidth]{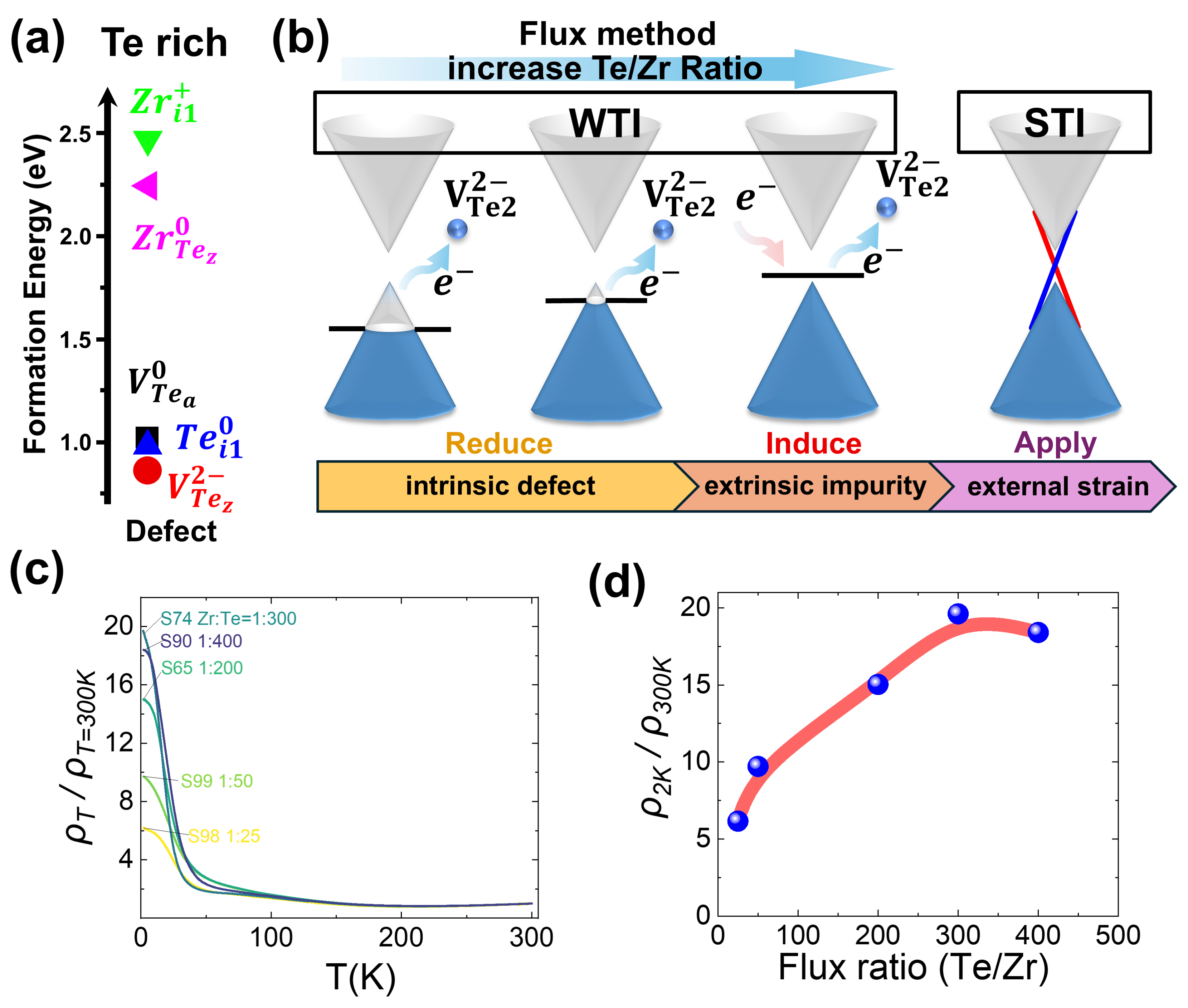}
	\caption{(a)  Formation energies of intrinsic point defects evaluated at the valence-band maximum (VBM) under Te-rich conditions. (b) Schematic illustration of electronic and topological phases in ZrTe$_{5}$, arising from variations in V$_{{Te}_z}$ defect concentration. The horizontal axis qualitatively increases the Te/Zr ratio. Ideal WTI induced by extrinsic impurity in lightly $\mathrm{V}_{\mathrm{Te_z}}^{2-}$ defect sample and STI upon applying external strain to the ideal WTI. (c),(d) Temperature-dependent resistivity $\rho_{xx}$ for flux-grown ZrTe$_{5}$ with different flux ratios. (c) Normalized temperature-dependent resistivity for ZrTe$_{5}$ samples grown with different flux ratios. S74 data from Ref.\cite{exp_rho}. (d) Variation of $\rho_{2K}$⁄$\rho$$_{300K}$  as a function of flux ratio.} \label{fig4}
\end{figure*}

Finally, we discuss how to control intrinsic defects during sample synthesis to achieve stable and well-defined electronic states in ZrTe$_5$. Notably, the formation energies of all intrinsic defects are positive in the Te-rich regime [Figure \ref{fig4}(a) and Figure S4], indicating that increasing the Te/Zr ratio effectively suppresses overall defect formation in ZrTe$_5$.  Given that the main defect in the Te-rich regime is $\mathrm{V}_{\mathrm{Te}_{\mathrm{z}}}^{2-}$, we could see that increasing the Te/Zr ratio could realize a nearly ideal WTI phase with very light hole pockets. Experimentally, such Te-rich conditions can be realized via the flux-growth method, which provides an excess of Te during crystal growth [Figure \ref{fig4}(b)]. Therefore, we expect the sample resistance to increase, reflecting a trend toward the ideal WTI phase with reduced hole pockets, when the Te/Zr ratio increases in the flux method.

Indeed, in the previous report, most of the observations in flux-grown samples are associated with hole-doped cases~\cite{FL2,FL-h1,FL4,FL6,FL7}. To further verify our predictions, we synthesized five flux-grown samples with Zr/Te ratios of 1:25, 1:50, 1:200, 1:300, and 1:400 and measured their resistivity. Figure \ref{fig4}(b) displays the normalized resistivity-temperature ($R$-$T$) curves, which show a systematic enhancement of low-temperature resistivity with increasing Te content. This trend is quantified in Figure \ref{fig4}(c): the $\rho_{2K}/\rho_{300K}$ ratio increases monotonically with Te/Zr ratio. These results confirm our theoretical prediction that enhancing Te richness effectively suppresses defect formation and drives ZrTe$_5$ toward a nearly ideal weak topological insulator state.

Having established an approach to realize nearly ideal WTI phases in ZrTe$_5$ through intrinsic defect control, we note that even under highly Te-rich flux-growth conditions, where intrinsic defects are greatly suppressed, achieving a completely defect-free crystal is practically impossible. Consequently, slight hole pockets remain unavoidable in samples with very high Te/Zr ratios. To obtain a truly ideal topological phase, where the Fermi level lies precisely within the bulk gap, extrinsic impurities must be introduced [Figure \ref{fig4}(b)]. In particular, these extrinsic dopants should donate a small number of electrons to compensate for the residual hole pockets caused by $\mathrm{V}_{\mathrm{Te_z}}^{2-}$ defects.  For example, incorporating iodine — which has one more valence electron than Te — into the flux during crystal growth under Te-rich conditions allows iodine to substitute for a small number of Te sites \cite{Idoped}, thereby pushing the Fermi level into the gap. However, realizing the ideal WTI requires fine-tuning of extrinsic impurities and warrants further investigation in future experimental studies.


Our calculations further indicate that realizing the STI phase requires a relatively high density of intrinsic defects. To fine-tune the Fermi level exactly into the bulk gap, precise control over the relative concentrations of Zr interstitials and Te vacancies at the zigzag sites is necessary, an experimentally challenging task. Therefore, we propose that a nearly ideal STI phase may instead be achieved by applying external strain to samples synthesized under high Te/Zr flux ratios [Figure \ref{fig4}(b)]~\cite{FL6}.

 In summary, our work establishes a clear and practical strategy for preparing ZrTe$_5$ samples with stabilized and well-defined electronic states, which is essential for advancing both fundamental studies and device applications of this promising quantum material. Beyond enabling the realization of ideal topological phases in ZrTe$_5$, our approach also facilitates the exploration of other intriguing states that rely on controlled bulk electronic properties. For example, our theoretical framework may pave the way for achieving the ideal three-dimensional quantum Hall effect in ZrTe$_5$, where appropriate electron filling of bulk states is required \cite{mobility}.


\medskip

%


\clearpage

\clearpage

\end{document}